\documentclass[reprint,aps,prl,twocolumn,superscriptaddress,preprintnumbers,letterpaper,natbib,floatfix]{revtex4-1}
\usepackage{amssymb}
\usepackage{amsmath}
\usepackage{epsfig}
\usepackage{comment}
\usepackage{color}
\usepackage{hyperref}
\usepackage{cleveref}
\usepackage{multirow}
\usepackage{capt-of}
\usepackage{siunitx}
\usepackage{graphicx}
\usepackage{slashed}
\usepackage{tabularx}
\usepackage{enumitem}
\usepackage{multirow}
\usepackage{booktabs}
\usepackage{isotope}
\usepackage{xcolor}
\usepackage{bm}
\usepackage{array,mathtools}

\newcommand{\mysec}[1]{\paragraph*{#1.}\!\!\!\!\!---}

\newcommand{\bea}{\begin{eqnarray}}
	\newcommand{\eea}{\end{eqnarray}}

\newcommand{\psiB}{\psi_{\mathcal{B}}}
\newcommand{\phiB}{\phi_{\mathcal{B}}}
\newcommand{\mpsiB}{m_{\psi_{\mathcal{B}}}}
\newcommand{\mphiB}{m_{\phi_{\mathcal{B}}}}
\newcommand{\yd}{y_d}
\newcommand{\GE}[1]{{\color{red} [GE: #1] }}

\begin{document}
	\title{Dark Matter Induced  Nucleon Decay Signals in Mesogenesis}
	\author{Joshua Berger}
	\email{Joshua.Berger@colostate.edu}
	\affiliation{Colorado State University, Fort Collins, Colorado 80523, USA}
	\author{Gilly Elor}
	\email{gilly.elor@utexas.austin.edu}
	\affiliation{Weinberg Institute, Department of Physics, University of Texas at Austin, Austin, TX 78712, USA}
	
	\begin{abstract}
		We introduce and study the first class of signals that can probe the dark matter in Mesogenesis which will be observable at current and upcoming large volume neutrino experiments. The well-motivated Mesogenesis scenario for generating the observed matter-anti-matter asymmetry necessarily has dark matter charged under baryon number. Interactions of these particles with nuclei can induce nucleon decay with kinematics differing from sponanteous nucleon decay. We calculate the rate for this process and develop a simulation of the signal that includes important distortions due to nuclear effects. We estimate the sensitivity of DUNE, Super-Kamiokande, Hyper-Kamiokande, and JUNO to this striking signal.
	\end{abstract}
	
	\maketitle

	Mesogenesis mechanisms \cite{Elor:2018twp,Elor:2020tkc,Elahi:2021jia,Alonso-Alvarez:2019fym} utilize the Charge-Parity (CP) violation of Standard Model (SM) meson systems to generate the primordial baryon asymmetry and the dark matter (DM) abundance of the Universe. Excitingly,  Mesogenesis is highly testable \cite{Elor:2022jxy,Alonso-Alvarez:2021qfd,Elor:2022jxy} and
	experimental searches are underway to probe signals directly linked to the generated baryon asymmetry \cite{Belle:2021gmc, Rodriguez:2021urv,Alonso-Alvarez:2021qfd, Borsato:2021aum,Shi:2022rjn} (see overviews in \cite{Elor:2022hpa,Asadi:2022njl,Barrow:2022gsu,Goudzovski:2022vbt}). However,  a direct probe of the DM in Mesogenesis remained elusive, until now. In this Letter we study DM induced nucleon decays (IND) in Mesogenesis which can produce striking signals at current and upcoming neutrino detectors.
	While the Mesogenesis framework is the focus of this letter, methods developed here can be broadly applied to search for models containing dark baryons e.g. \cite{Alonso-Alvarez:2021oaj, Fornal:2020gto, Grinstein:2018ptl}.
	
	The novel way in which Mesogenesis satisfies the Sakharov conditions \cite{sakharov},
	is as follows: mesons produced at late (MeV scale) times, having undergone CP violating processes, decay out of equilibrium into dark baryons. This process generates an equal and opposite baryon asymmetry between the dark and visible sector \emph{e.g.} in $B^0$-Mesogenesis \cite{Elor:2018twp} the baryon asymmetry is generated from the late time production of $B_{s,d}^0$ mesons which undergo CP violating oscillations before quickly decaying into a SM baryon and a dark Dirac fermion $\psiB$ carrying SM baryon number $-1$.
	To evade washing out the generated baryon asymmetry through e.g. $\psiB \rightarrow p \,e \,\bar{\nu}_e$, the $\psiB$s must rapidly decay into stable DM states.
	
	Mesogenesis DM consists of a dark Majorana fermion $\xi$, and a dark complex scalar $\phiB$ which is charged under SM baryon number. These two stable  particles compose the entirety of DM \emph{i.e.} the DM halo will consist of a mixture of $\xi$ and $\phiB$ which can scatter off target nuclei in neutrino detectors to produce mono-energetic mesons and missing energy. This process of IND appears experimentally as nucleon decay but with different kinematics, as such current limits are not constraining.
	
	The cross section for IND, as it arises in Mesogenesis, will be within reach of neutrino detectors and can be searched for at the Deep Underground Neutrino Experiment (DUNE) \cite{Guinot:2022smb}, Super-Kamiokande \cite{Walter:2008ys}, Hyper-Kamiokande \cite{Yokoyama:2017mnt}, and JUNO \cite{JUNO:2015sjr}. Furthermore, given the mono-energetic (up to smearing effects) meson, such signals should be distinct over SM backgrounds which primarily due to atmospheric neutrino processes.
	
	To study the details of the IND process, we developed a Monte Carlo event generation tool within the \texttt{GENIE}~\cite{Andreopoulos:2009rq,Andreopoulos:2015wxa} software suite used to study neutrino scattering events \footnote{All data generated using this code is available upon request. The code itself can be downloaded from https://github.com/jberger7/Generator-IND}.  We can then study the detailed kinematics of the outgoing mesons produced during DM scattering events and compare these with the dominant atmospheric neutrino scattering background. The event generation includes nuclear effects which smear out the spectrum of mesons from one that is nearly mono-energetic in the non-relativistic DM limit. Furthermore, these events are ready to be used for study in neutrino experiments, where they can be passed through detector-specific simulation software.
	
	In this Letter we first characterize the IND signal. We present the Mesogenesis specific parameter space which can be targeted by experiments. We then detail the simulation.  Next, we apply this simulation to estimate the sensitivity of experiments.

	\mysec{Characterizing the IND Signal}   Generating the baryon asymmetry in $B$-Mesogenesis \cite{Elor:2018twp,Elahi:2021jia}, requires the existence of a  TeV scale colored scalar \footnote{Which may be identified as a squark in a supersymmetric embedding \cite{Alonso-Alvarez:2019fym}.}  which mediates the baryon number conserving decay of a $B$ to $\psiB$ and a SM baryon through the (GeV scale) effective operator:
	\bea
	\label{eq:OpBdecay}
	&& \mathcal{O}_{ab,c} =  C_{ab,c} \epsilon_{ijk} \left(u_{a}^i d_{b}^j\right)\bigl(\psiB \, d_{c}^k\bigr)\, ,
	\eea
	where all fermions are right handed, though our results generalize to left-handed operators.  $C_{u_a d_b, d_c} \equiv y_{u_a d_b} y_{\psi d_c} / M_Y^2$, where $M_Y$ is the mediator mass. $\psiB$ decays to the DM through $\mathcal{L}_d \supset \yd \bar{\psi}_{\mathcal{B}} \xi \phiB + \text{h.c.}$, which is allowed by a stabilizing $\mathbb{Z}_2$ symmetry under which $\psiB$ is even and the DM particles odd. $\yd$ is a free parameter, but  a motivated benchmark is $\yd \lesssim \mathcal{O}(0.1)$ which results in the correct DM abundance given an example UV embedding \cite{Alonso-Alvarez:2019fym}.
	
	\begin{figure}[t!]
		\centering
		\includegraphics[width=0.8 \columnwidth]{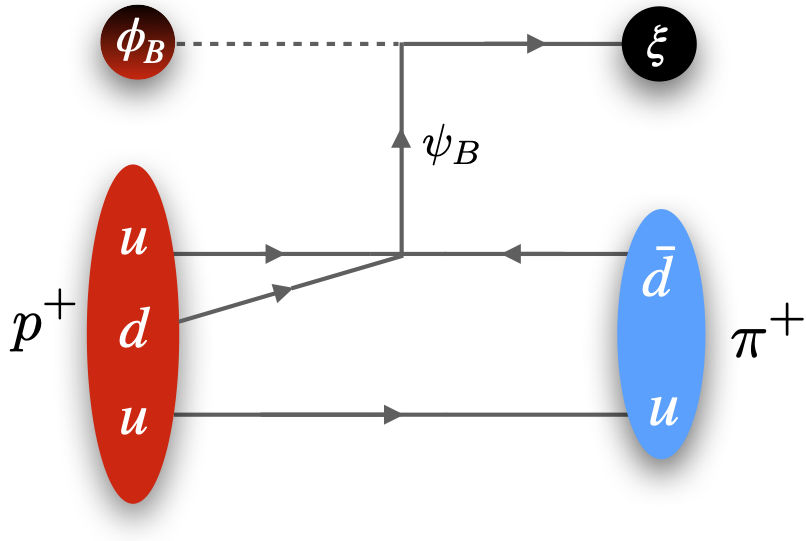}
		\caption{Induced proton decay to a pion through $\mathcal{O}_{u,dd}$.}
		\label{fig:cartoon}
	\end{figure}

	Eq.~\ref{eq:OpBdecay} generates the IND signal in Mesogenesis: when kinematically allowed, an incoming $\xi$ or $\phiB$ scatters off a proton or neutron by exchanging a $\psiB$ and produces an energetic meson. Fig.~\ref{fig:cartoon} depicts an example process ---  incoming $\phiB$'s induce proton decay to $\pi^+$ through $\mathcal{O}_{ud,d}$. Similarly, IND to kaons arises through $\mathcal{O}_{ud,s}$ and $\mathcal{O}_{us,d}$.  We consider searches for the processes:
	\begin{subequations}\label{eq:InducedNucDecay}
		\begin{align}
			& \phiB \, N \rightarrow \mathcal{M} \,  \xi \,\,\,\,\, \text{if} \,\,\,\,\, \mphiB + m_N  >  m_{\mathcal{M}} + m_\xi \,, \label{eq:phidecay}\\
			& \xi  \,  N \rightarrow \mathcal{M}  \,  \phiB^\star \,\,\,\,\, \text{if} \,\,\,\,\, m_{\xi} + m_N >  m_{\mathcal{M}} +  \mphiB\,, \label{eq:xidecay}
		\end{align}
	\end{subequations}
	where $N = n^0$, $p^+$ and $\mathcal{M}$ is a SM meson. Recall that $\xi$ is Majorana, allowing any of the DM states to participate in this process when it is kinematically allowed.
	For decays induced by incoming $\phi$s, the kinetic energy of the outgoing meson, to $ \mathcal{O}(v_{\text{DM}})$, is given by
	\bea
	\label{eq:MER}
	E^{\mathcal{M},\, \text{kin}}_{\phiB N \rightarrow \xi \mathcal{M}} = \frac{m_{\mathcal{M}}^2 - m_{\xi}^2 + \left( m_N + m_{\phi_B} \right)^2}{2 (m_N + m_{\phi_B})} - m_{\mathcal{M}}\,.
	\eea
	Swap $m_\xi \leftrightarrow \mphiB$ to obtain the meson energy from incoming $\xi$s.

	If the  struck nucleon is at rest, then the outgoing meson is mono-energetic with energy given in Eq~\eqref{eq:MER}. 
	However, the nucleons are moving with a momentum of $\mathcal{O}(100~\text{MeV})$ inside nuclei, smearing out of the meson signal (except for the case of scattering off hydrogen in water Cherenkov detectors). We simulate the IND process, carefully accounting for this smearing. Note that the energies of these decays are shifted compared to spontaneous nucleon decay, with higher energies when $\phiB$ scatters and lower energies when $\xi$ scatters. This alters the phenomenology of the Mesogenesis scenario compared with proton decay models such as grand unified theories, the targets of current nucleon decay searches \cite{McGrew:1999nd, Soudan2:2000gbz,Super-Kamiokande:2013rwg,Super-Kamiokande:2016exg, Super-Kamiokande:2005lev,Super-Kamiokande:2014otb}.  As such, existing limits from nucleon decay searches do not generally constrain the Mesogeneis signal \footnote{similar considerations were discussed in the context of Hylogenesis \cite{Davoudiasl:2011fj}}. 
	
	The cross section for IND is obtained from the matrix element:
	\bea
	\label{eq:MatrixElement}
	\mathcal{A}_{\phiB N \rightarrow \xi \mathcal{M}} & = \,\, \bar{u}_\xi (\vec{p}_\xi )\frac{\yd \, C_{ab,c} }{p_{\psi_B}^2 - m_{\psi_B}^2} \left(\slashed{p}_{\psi_B} + m_{\psi_B} \right) \\ \nonumber
	&\quad \times \, P_{R} \Bigl[ W_0^{RR} - i \, \frac{\slashed{p}_{\psi_B}}{m_N}  W_1^{RR} \Bigr] u_N (\vec{p}_N)\,,
	\eea
	with $\bar{u}_\xi \rightarrow \bar{v}_\xi$ for $\xi \, N \rightarrow \mathcal{M}\, \phiB^\star $. Here, $p_{\psi_B} = p_\xi - p_{\phi_B}$ and $P_{R}$ 
	is the right handed fermion projector. 
	The Wilson Coefficients are constrained by a combination of LHC searches for the mediator and flavor observables:  $C_{ud,d}^{\text{max}} = 0.07 \,\text{TeV}^{-2}$ and $C_{ud,s}^{\text{max}} \,, \, C_{us,d}^{\text{max}} = 0.64  \,\text{TeV}^{-2}$ \cite{Alonso-Alvarez:2021oaj, Alonso-Alvarez:2021qfd}.
	Since INDs can lead to $\mathcal{O}(\text{GeV})$ momentum transfer, we use high $q^2$ extrapolated lattice results for the form factors $W_{0,1}^{RR}$  \cite{Aoki:2017puj}  \footnote{The corresponding values of  $W_{0} (W_{1})$ for $\pi^+$ are  $0.038\sqrt{2}$ ($-0.05\sqrt{2}$), for $\pi^0$ are 0.038 (-0.05), and, through $\mathcal{O}_{ud,s}$, for $K^+$ are 0.0648 (-0.046) and for $K^0$ are -0.0648 (0.046).}. Including the momentum dependence of $W_{1,2}$ negligibly affects the signal. $\mathcal{O}_{us,d}$ and $\mathcal{O}_{ud,s}$ require different form factors but lead to similar signals.

					\begin{table}[t!]
\renewcommand{\arraystretch}{1.2}
  \setlength{\arrayrulewidth}{.25mm}
\centering
\small
\setlength{\tabcolsep}{0.18 em}
\begin{tabular}{ | c  | c | c | c | }
    \hline
   Initial						&  Final   	&  Meson 		&  Approx. $ \langle \sigma v \rangle_0$    \\
      DM 					& Meson& $E_{\rm Kin}$ [GeV] 	&	$ \left[ \text{cm}^3 / \text{sec} \right]$	     \\ \hline   \hline
$\phiB$  	 &  $\pi^+/\pi^0$  	&	  0.6 - 1.2  	& $10^{-21.4}$ - $10^{-21.0}$ \\ \hline
$\xi$  	 &  $\pi^+/\pi^0$  & 	  0.02 - 0.6     & $10^{-22.5}$ - $10^{-21.9}$ \\ \hline
$\phiB$  	 &  $K^+/K^0$   & 	  0.3 - 0.9 	& $10^{-19.7}$ - $10^{-19.3}$  \\ \hline
$\xi$  	 &  $K^+/K^0$   & 	 0.04 - 0.3  	& $10^{-20.6}$ - $10^{-19.8}$.\\ \hline
\end{tabular}
\caption{ 
IND allowed by Eqs.~\eqref{eq:InducedNucDecay} and \eqref{eq:Constraints}, the expected range of un-smeared kinetic energy of the outgoing meson and the stripped cross section defined in Eq.~\eqref{eq:sigma0def}.}
\label{tab:IND}
\end{table}
	
\begin{figure*}[t]
		\centering
		\setlength{\tabcolsep}{0pt}
		\renewcommand{\arraystretch}{0}
		\begin{tabular}{cccc}
			\label{fig:EkinPHIpnPI}
			\includegraphics[width=0.24\textwidth]{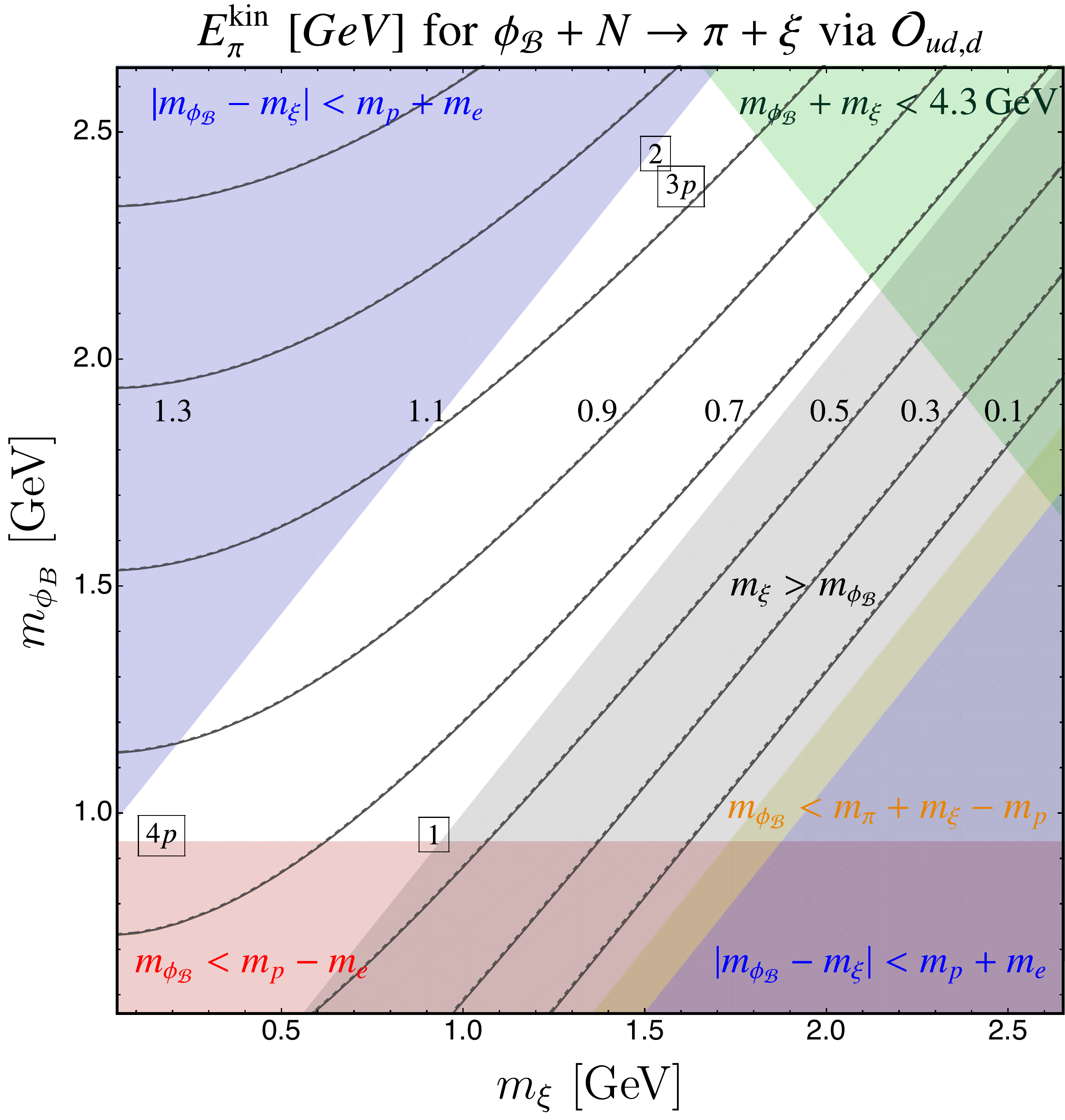}
			
			\label{fig:EkinXIpnPI}
			\includegraphics[width=0.24\textwidth]{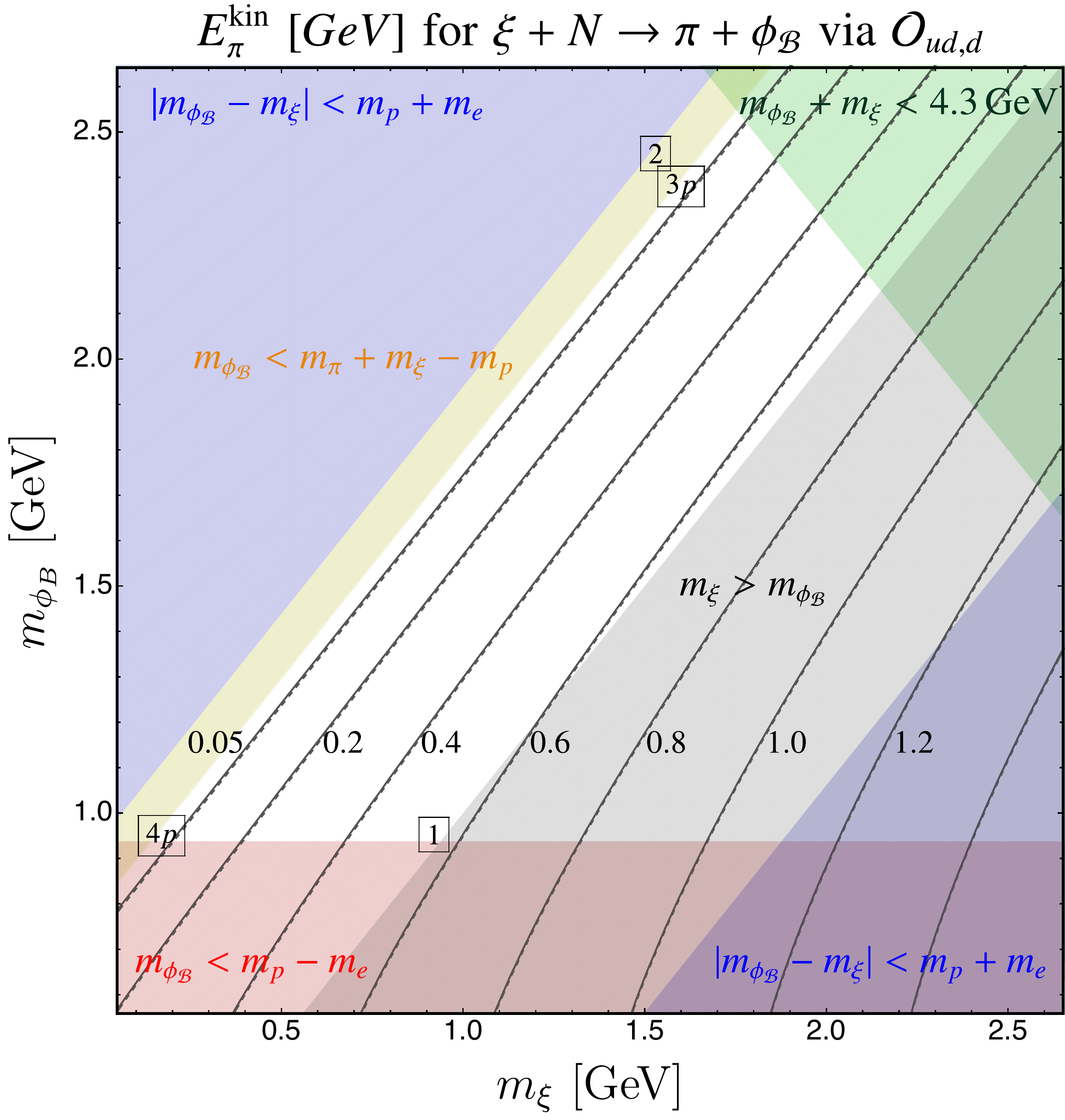}
			
			\label{fig:EkinPHIpnKuds}
			\includegraphics[width=0.24\textwidth]{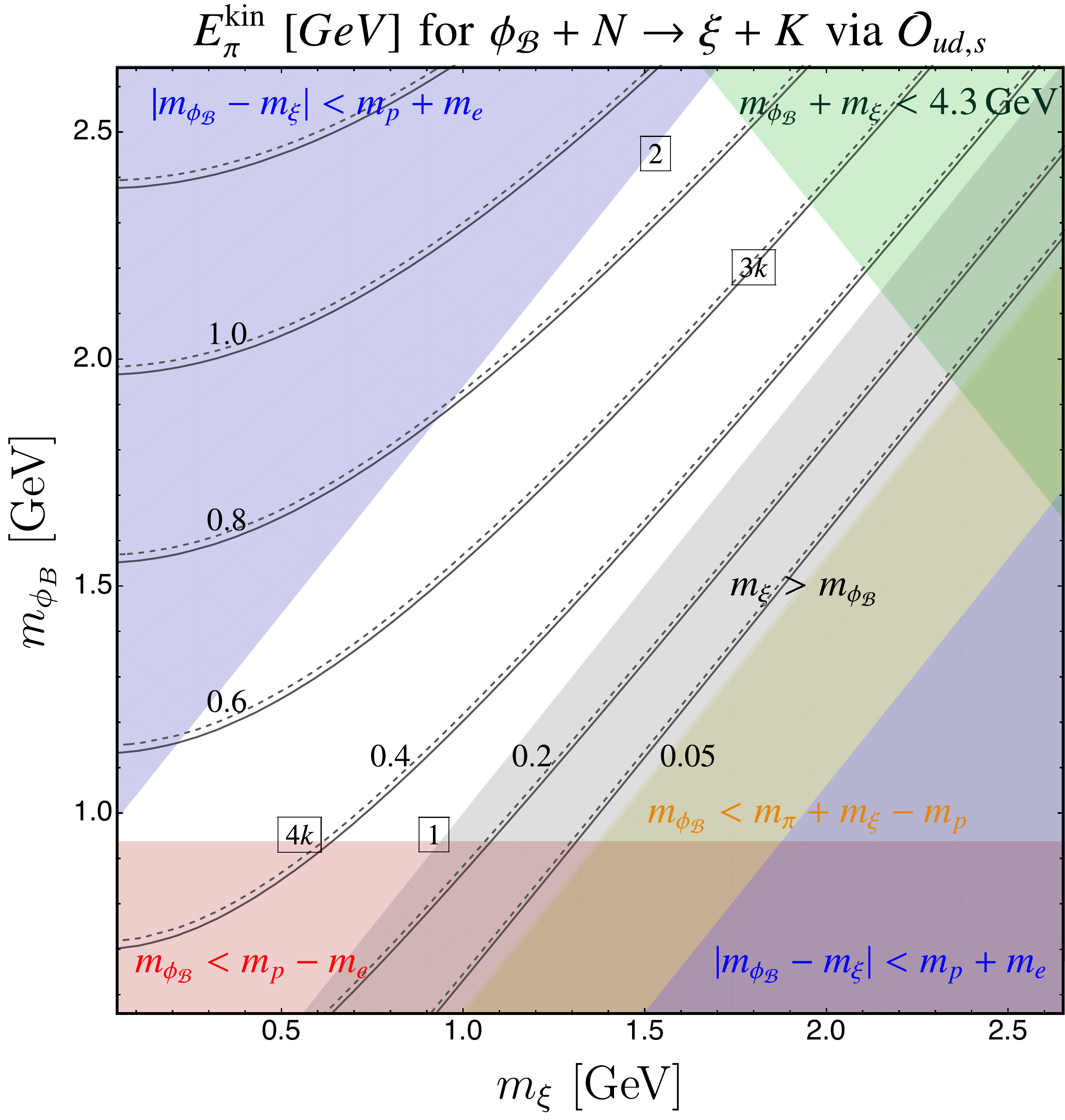}
			
			\label{fig:EkinXIpnKuds}
			\includegraphics[width=0.24\textwidth]{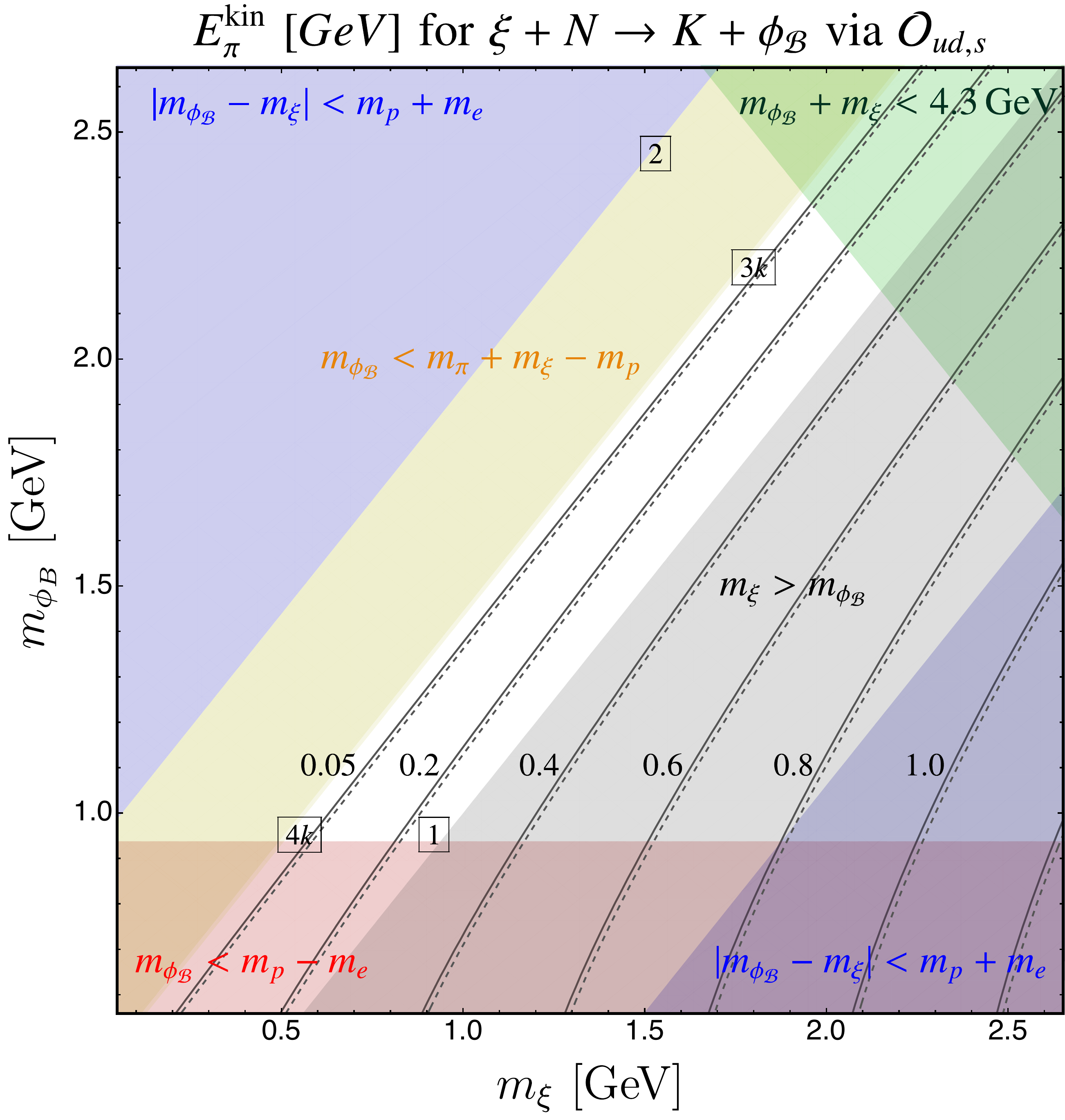}
			
		\end{tabular}
		\vspace{-0.2cm}
		\caption{Parameter space and kinetic energy contours for the eight different DM IND proceses arising in Mesogenesis.  Colored regions are ruled out by kinematics  Eq.~\eqref{eq:InducedNucDecay} or mechanism considerations Eq.~\eqref{eq:Constraints}. Sold lines correspond to kinetic energy for scattering off protons $N=p$, and the dashed off nucleons $N=n$.
			In each panel, we indicate the location of the representative BMs: 1, 2, 3p, 4p for pions and 1, 2, 3k, 4k for kaons as summarized in Table ~\ref{tab:BMs}.}
		\label{fig:MasterParamPlot}
	\end{figure*}

	\mysec{Mesogenesis Parameter Space} In addition to the kinematic constraints on IND Eq.~\eqref{eq:InducedNucDecay}, the allowed parameter space is constrained by Mesogenesis-specific considerations \emph{i.e.} generating the observed baryon asymmetry and DM abundance:
	\begin{subequations}
		\label{eq:Constraints}
		\begin{align}
			& \mphiB + m_\xi < \mpsiB < m_B - m_p \simeq 4.34 \, \text{GeV}\,, \label{eq:psiBm}  \\
			& |m_{\phiB} - m_\xi | < m_p + m_e \simeq 938.8 \, \text{MeV}\,,\label{eq:noCoales} \\
			& \mpsiB \,,\, \mphiB > m_p - m_e\,. \label{eq:mphiMin} \\
			& \mphiB > m_\xi  \,. \label{eq:DMconstraint}
		\end{align}
	\end{subequations}
	The regions in $\{m_\xi, \mphiB\}$ space excluded by these constraints are shaded in Fig.~\ref{fig:MasterParamPlot}, while the white region is allowed.
	Eq.~\eqref{eq:psiBm} ensured that $\psiB$ is light so that the baryon asymmetry can be generated through the decay $B \rightarrow \mathcal{B}_{\rm SM} + \psiB$, while also being heavy enough to decay into the DM $\xi$ and $\phiB$. Decreasing the value of $\mpsiB$ corresponds to increasing the excluded green region. For  $\mpsiB \simeq 1.1$ GeV, there is no longer viable parameter space. Eq.~\eqref{eq:noCoales} is enforced to prevent $\xi$ and $\phiB$ from coalescing into SM baryons, which would washout the asymmetry.  Proton decay through Eq.~\eqref{eq:OpBdecay} to dark baryons is kinematically forbidden through Eq.~\eqref{eq:mphiMin}.

	There exist dark sector interactions which deplete the DM and ensure the correct abundance \cite{Elor:2018twp}.  Since $n_{\phi_{\mathcal{B}}} - \bar{n}_{\phi_{\mathcal{B}}}$ is related to the baryon asymmetry, $\phiB$ must always constitute some, but not all, of the DM: if $\mphiB < m_\xi$, dark interactions could annihilate the entire $\xi$ population into $\psiB$s. The measured ratio of DM to baryon densities \cite{Planck:2018vyg} implies that
	$\mphiB \simeq 5 m_p$, violating Eq.~\eqref{eq:psiBm}.
	This motivates multi-component DM where we enforce Eq.~\ref{eq:DMconstraint} so that the symmetric component of $\phiB$s annihilate into $\xi$s.
	Since the measured SM baryon asymmetry is always balanced by an asymmetry in $\phiB$s, the observed DM to baryon ratio $\rho_{\rm DM} \sim 5 \rho_{\mathcal{B}}$ fixes the expected density of $\xi$ and $\phiB$ particles in the halo: $\rho_\xi/\rho_{\phiB} = 5 m_p/\mphiB - 1$, and $\rho_{\rm total} = \rho_\phi + \rho_\xi = 0.4\, \text{GeV}/\text{cm}^3$. Given Eq.~\eqref{eq:mphiMin},  there will always be a substantial asymmetric component of DM, and so
	both INDs in Eq.~\eqref{eq:InducedNucDecay} will be present if kinematically allowed.

	The scattering cross sections for the INDs are computed from Eq.~\eqref{eq:MatrixElement}.
We parametrize the cross section as:
	\bea
	\label{eq:sigma0def}
	\langle \sigma v \rangle_{\rm DM}^{\mathcal{M}}  \equiv \frac{(\yd \times C_{u d_i,d_j} )^2}{\mpsiB^2 [\text{GeV}^{-4}]} \langle \sigma v \rangle_{\rm DM}^{\mathcal{M},0}
	\eea
	where $\mathcal{M} = \pi^0$, $\pi^+$, $K^+$, $K^0$ and $\text{DM} = \phiB$, $\xi$. The range of variation $\mpsiB \sim $1-4.3 GeV is a small effect. 
	Values of the coupling stripped cross section $\langle \sigma v \rangle_{\rm DM}^{\mathcal{M},0}$  over the allowed parameter space of Fig.~\ref{fig:MasterParamPlot} are shown in Table~\ref{tab:IND}. For e.g. $\yd = 0.1$ and $C^{\rm max}_{u d_i,d_j}$, the expected cross section can be as large as $10^{-38} - 10^{-36} \text{cm}^3/\text{sec}$ in the allowed parameter space for all channels. Meanwhile, the estimated detector sensitivity is $\sim10^{-42}-10^{-40} \text{cm}^2/\text{sec}$ .
	
	\setlength{\tabcolsep}{0.24 em}
	\begin{table}[t!]
		\renewcommand{\arraystretch}{1.1}
		\setlength{\arrayrulewidth}{.25mm}
		\centering
		\begin{tabular}{| c || c | c | }
			\hline
			Benchmark & $\mphiB $ [GeV] & $m_\xi$  [GeV]    \\
			\hline \hline
			1 & 0.95  & 0.92    \\
			2 & 2.45  & 1.53       \\
			3p & 2.38  & 1.6   \\
			3k & 2.2  &  1.8       \\
			4p & 0.95  &  0.17  \\
			4k & 0.95  &  0.55  \\
			\hline
		\end{tabular}
		\caption{BMs highlighting the possible signal topologies. BM 1 corresponds to $E^{\rm kin, min}_{p \phiB \rightarrow \mathcal{M} \xi } \sim E^{\rm kin, max}_{p \phiB \rightarrow \mathcal{M} \xi }$ for both $\pi$s and kaons. BM 2 corresponds to $E^{\rm kin, max}_{p \phiB \rightarrow \mathcal{M} \xi }$.  BMs 3p and 3k correspond to the maximal $E^{\rm kin}_{p \phiB \rightarrow \mathcal{M} \xi }$ such that the incoming $\xi$ process is still allowed for production of $\pi$s and kaons respectfully. BMs 4p and 4k highlight a region of $\{m_\xi, \mphiB \}$ which would still lead to a signal for small $\mpsiB$ (see labels in Fig~\ref{fig:MasterParamPlot}).
		}
		\label{tab:BMs}
	\end{table}
	
	\mysec{Signal Monte Carlo and Benchmarks}
	We pick benchmark (BM) points to highlight the possible signal topologies; these are defined in Table~\ref{tab:BMs} and labeled in Fig.~\ref{fig:MasterParamPlot}. For large nuclei, nuclear effects including motion of the nucleons and final state interactions of hadronic particles escaping the nuclear remnant smear the outgoing meson energy, can liberate additional hadrons, and can change the isospin characteristics of the meson. In order to account for these effects, as well as allow for future simulation of the detailed detector response, we have developed a Monte Carlo event generation tool for the IND process.
	
	Signal events are generated using a modified version of \texttt{GENIE v3.0.2}
	~\cite{Andreopoulos:2009rq,Andreopoulos:2015wxa}. We employ the default tune (\texttt{G18\_02a}) throughout, though we considered other nuclear models. We found differences in the signal distributions of order 10\% between hA and hN models of the intranuclear cascade. The current nucleon decay module in \texttt{GENIE} was modified to allow for IND kinematics. The meson final states currently implemented are $\pi^0$, $\pi^+$, $K^0_{S/L}$, $K^+$, and $D^0$. This module propagates the outgoing mesons through the nuclear remnant to the edge of the nucleus. The kinematics of the IND process are fixed given masses for the two DM particles. The cross section is determined by Eq.~\eqref{eq:sigma0def}. For non-relativistic DM, there is a small difference in the rate for interaction with a high speed nucleon compared with a nucleon at rest. We neglect this small difference of $\sim$10\%, as well as the nuclear modeling uncertainties, in our sensitivities below.

	\mysec{Signals at Neutrino Experiments}
	The current best limits on spontaneous nucleon decays to pions are from Super-Kamiokande  \cite{Super-Kamiokande:2013rwg},
	which applied a pion momentum cut of 1 GeV and are thus applicable to parts of Mesogenesis parameter space --- here the experimental limit can be compared to an effective lifetime $(\tau_N^{\text{Ind}})^{-1} = n_{\rm DM} \langle \sigma v \rangle_{\rm DM}$ \cite{Davoudiasl:2010am, Davoudiasl:2011fj,Huang:2013xfa}.
	This yields an approximate, conservative, limit of  $\yd \times C_{ud,d}/C_{ud,d}^{\rm max} \gtrsim 0.03-0.1$ for $70 \,  \text{MeV} \lesssim E_{\pi^+}^{\rm kin} \lesssim 870 \, \text{MeV}$ and $E_{\pi^0}^{\rm kin} \lesssim 870 \, \text{MeV}$. Existing searches in kaon channels \cite{Super-Kamiokande:2005lev, Super-Kamiokande:2014otb,JUNO:2022qgr} placed narrow ranges of momentum cuts. Consequentially, the majority of parameter space of interest  is unconstrained by existing searches. Super-Kamiokande is expected to have sensitivity to IND signals given \emph{dedicated} studies.  Hyper-Kamiokande will improve on this with its larger exposure.  Since DUNE is based on liquid argon time-projection chamber (LArTPC) technology, it could have particular sensitivity to certain models.  Liquid Scintillators, e.g. JUNO, have low thresholds and we expect high sensitivity particularly to charged particles. Our modeling of the different detector
responses is described in the Supplementary Material.

		\begin{figure*}[t]
		\centering
		\setlength{\tabcolsep}{4pt}
		\renewcommand{\arraystretch}{1}
		\begin{tabular}{cc}
			\includegraphics[width=0.45\textwidth]{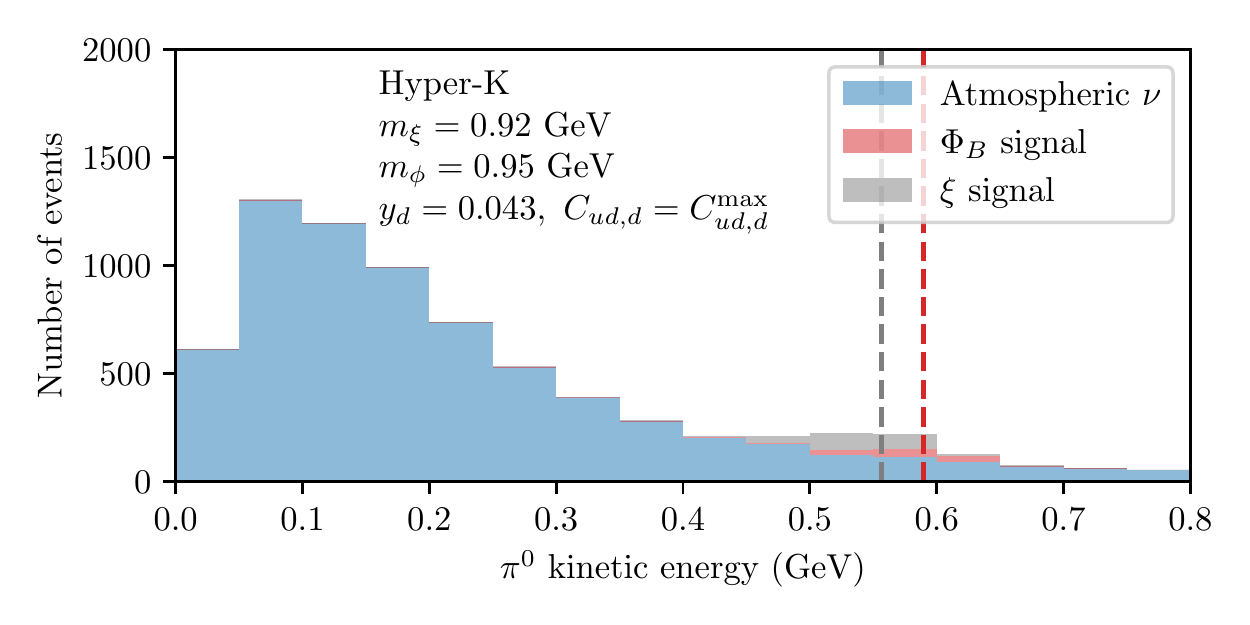}
			&
			\includegraphics[width=0.45\textwidth]{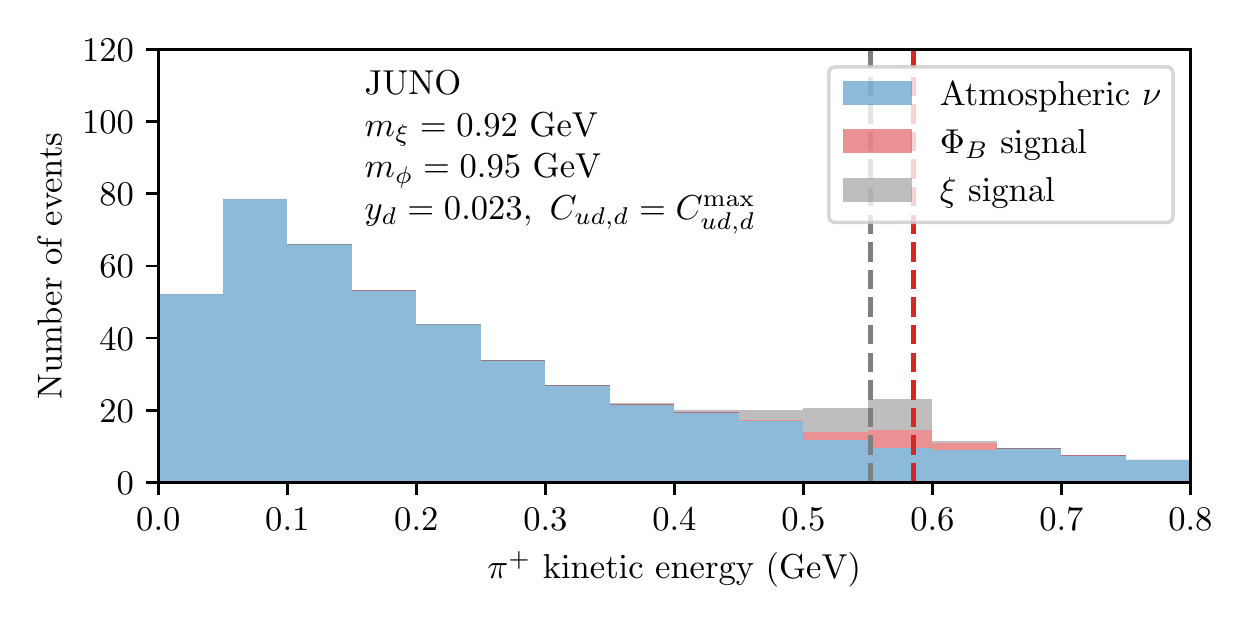} \\
			
		\end{tabular}
		\vspace{-0.3cm}
		\caption{ 
		Kinetic energy distributions for a sample BM models at Hyper-Kamiokande, and JUNO. Super-Kamiokande, is simply a rescaling of the rate at Hyper-Kamiokande by a factor of 16. Dashed lines indicate the un-smeared energy Eq.~\ref{eq:MER}. Green lines indicate, where relevant, the assumed threshold for the detector to see the meson. The Hyper-Kamiokande and JUNO signals have a mono-energetic spike corresponding to scattering of hydrogen, while the smeared distribution corresponds to scatterings off oxygen.  The couplings are chosen at our estimated threshold of sensitivity.
		}
		\label{fig:BMsignal}
	\end{figure*}

	\mysec{Sensitivity Estimates}
	The dominant background is expected to be inelastic scattering of atmospheric neutrinos off nuclei, leading to additional mesons. By looking for off beam timing events, beam-related backgrounds can be evaded. To study the atmospheric neutrino background, we generate atmospheric neutrino events using \texttt{GENIE v3.0.2}, along with the Bartol atmospheric neutrino flux model~\cite{Barr:2004br} at Soudan for DUNE and Kamioka or all other experiments.
	
\begin{figure}[t]
		\centering
		\setlength{\tabcolsep}{4pt}
		\renewcommand{\arraystretch}{1}
		\begin{tabular}{c}
			\includegraphics[width=0.45\textwidth]{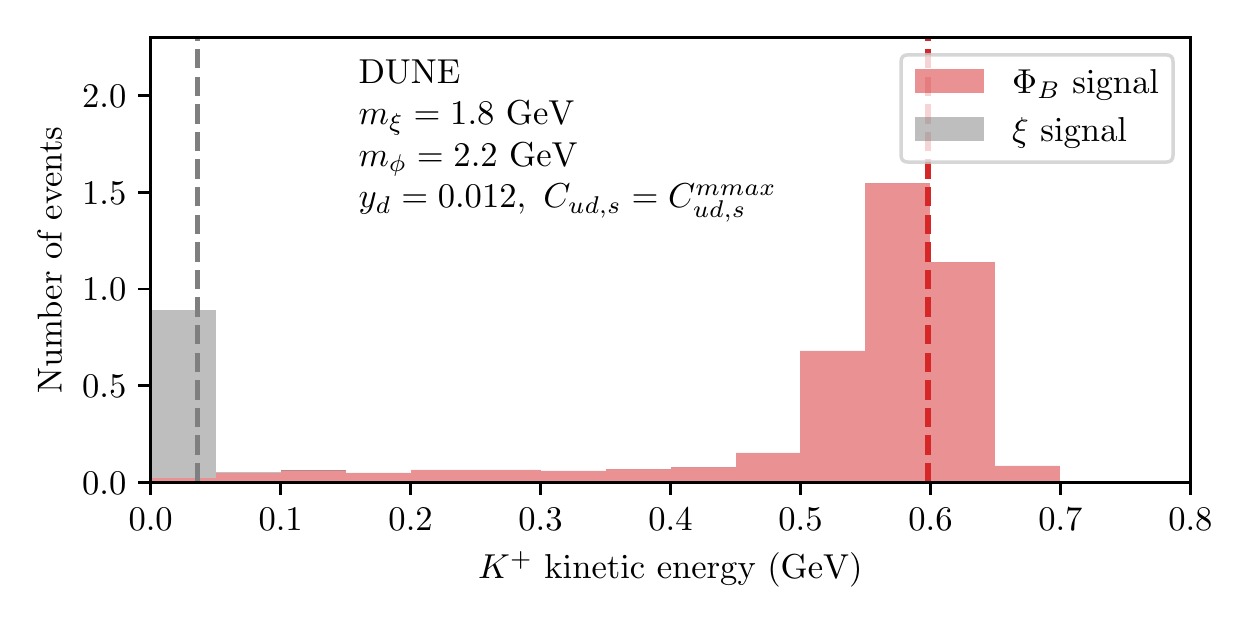} \\			
		\end{tabular}
		\vspace{-0.3cm}
		\caption{Kinetic energy distributions for sample BM models at DUNE for the a Kaon channel with the conventions outlined in Fig.\ \ref{fig:BMsignal}.		}
		\label{fig:BMsignalK}
	\end{figure}

	From samples of DM signals and atmospheric neutrino events, we look for events containing the relevant final state meson for the channel considered. Not all events with such a meson should be considered. Signal events contain a single meson and no other activity other than possible emissions from the nuclear remnant or byproducts of final state interactions. Thus, it is highly beneficial to veto any activity beyond the expected meson. To do so, we first apply thresholds (described in detail in the Supplemental Material \footnote{	
	Supplemental Material contains references \cite{Bueno:2007um,JUNO:2021tll,Esquivel:2018fls}, not already included in the text.} to all final state particles. Of the remaining particles that can be detected, we veto on events that have anything other than a meson of the expected type. For the pion channels, order 1\% of atmospheric neutrino scattering events lead to an event matching these criteria. This leaves a search that is not entirely background free. For the kaon channels on the other hand, single kaon events are only possible with additional flavor-violating weak interactions. Searches in these channels may thus be background free if kaon reconstruction is sufficiently good. To get a sense of the events after these vetoes we plot kinetic energy distributions of the remaining signal and background events; a few illustrative distributions are shown in Figs.~\ref{fig:BMsignal} and ~\ref{fig:BMsignalK}.
	
	We now estimate the sensitivity to $\yd \times C_{ud_i,d_j} / C_{ud_i,d_j}^{\rm max}$. For pion channels, we apply the selection described above.  To eliminate a majority of the background events, we also require that the selected pion has a kinetic energy within $100~\text{MeV}$ of its unsmeared value Eq.~\eqref{eq:MER}.  There are significant uncertainties in both the background flux and cross sections.  We therefore assume a $30\%$ background normalization uncertainty and determine an estimated $2\sigma$ sensitivity for the pion channels \footnote{See e.g.\cite{Super-Kamiokande:2023ahc}--- a recent   atmospheric neutrino analysis assigns uncertainties of order  $20-25\%$ on cross sections}.  For the kaon channels, no further selection is required as we have found that this channel is nearly background free, and we determine the coupling that would lead to 5 signal events over the assumed exposure of the experiment.  
	The sensitivity results are summarized in Table \ref{tab:sensitivity} for the BMs listed in Table~\ref{tab:BMs}.

	\mysec{Discussion}
	This Letter introduced the first direct probe of the DM in Mesogenesis; the  IND signal. We have chosen BMss i.e. Table~\ref{tab:BMs} that span the entire parameter space in which Mesogenesis is possible (as can be seen in Fig~\ref{fig:MasterParamPlot}). Therefore, an exhaustive search by the experiments discussed her could probe the entire parameter space of DM in Mesogenesis down to sensitivities listed in Table.~\ref{tab:sensitivity}. Furthermore, our sensitivity estimates in Table~\ref{tab:sensitivity} indicate that a null observation would place a more stringent bound on these operators than currently possible with colliders.  Figs.~\ref{fig:BMsignal}  and ~\ref{fig:BMsignalK} show the spectacular signal expected at DUNE, Hyper-Kamiokande, and JUNO for representative benchmark points. Note that DUNE is expected to have slightly better sensitivity to pion channels due to reduced backgrounds.  	
 In addition to providing experimentalists with the tools needed to search for DM IND signals at neutrino experiments, this letter paves the way to a signal driven model building effort of the dark sector. This work also paves the way to new the exploration of additional signals Mesogenesis through DM IND in high density astrophysical environments \footnote{Note that the effect of DM IND signals on Big Bang Nucleosynthesis (BBN) is expected to be small. $\rho_n/\rho_p$ will be negligibly affected as the cross section for the IND signal is nearly identical for protons and neutrons. Furthermore, $n_{DM} \langle \sigma v \rangle_{\rm IND}$ is significantly less than Hubble during the era of BBN when temperatures are of order a few MeV. Therefore, we expect the baryon to photon ratio to be negligibly effected.}, which is the subject of ongoing work.

		\setlength{\tabcolsep}{0.18 em}
	\begin{table}[t!]
		\renewcommand{\arraystretch}{1.1}
		\setlength{\arrayrulewidth}{.25mm}
		\centering
		\begin{tabular}{| c || c | c |  c | c |}
			\hline
			BM & DUNE& Super-K & Hyper-K& JUNO \\
			\hline  \hline
1 $\,\,\pi^+$ & 0.034 & 0.10 & 0.10 & 0.024 \\
2 $\,\,\pi^+$ & 0.015 & 0.054 & 0.054 & 0.011 \\
3p $\pi^+$ & 0.087 & 0.14 & 0.14 & 0.046 \\
4p $\pi^+$ & 0.11 & 0.18 & 0.17 & 0.045 \\ \hline
1 $\,\,\pi^0$ & 0.046 & 0.047 & 0.046 & 0.040 \\
2 $\,\,\pi^0$ & 0.023 & 0.021 & 0.019 & 0.020 \\
3p $\pi^0$ & 0.35 & 0.32 & 0.32 & 0.22 \\
4p $\pi^0$ & 0.13 & 0.12 & 0.12 & 0.084 \\ \hline
1 $\,\,K^+$ & 0.012 & 0.016 & 0.0070 & 0.019 \\
2 $\,\,K^+$ & 0.0077 & 0.0097 & 0.0042 & 0.011 \\
3k $K^+$ & 0.012 & 0.015 & 0.0065 & 0.017 \\
4k $K^+$ & 0.011 & 0.015 & 0.0062 & 0.017 \\ \hline
1 $\,\,K_S^0$ & 0.0024 & 0.0029 & 0.0012 & 0.0051 \\
2 $\,\,K_S^0$ & 0.0049 & 0.0058 & 0.0025 & 0.011 \\
3k $K_S^0$ & 0.0046 & 0.0054 & 0.0023 & 0.0097 \\
4k $K_S^0$ & 0.0029 & 0.0034 & 0.0015 & 0.0062 \\ 
			\hline
		\end{tabular}
		\caption{Estimated coupling sensitivity at DUNE with 400 kton-years of exposure, at Super-Kamiokande with 350.8 kton-years of exposure, at Hyper-Kamiokande with 1,900 kton-years of exposure, and at JUNO with 200 kton-years of exposure. We apply the solar minimum flux model for all experiments.  The solar maximum model gives slightly different sensitivity estimates.  All sensitivities are expressed as the ratio $ \yd  (C_{ud_i, d_j} / C_{ud_i,d_j}^{\rm max} )$; we have normalized the Wilson coefficients by the maximally allowed value from collider constraints. The fact that all values $\ll 1$, indicates that a IND searches at neutrino experiments are significantly more powerful probes than other existing experiments. }
		\label{tab:sensitivity}
	\end{table}

	\newpage
	\begin{acknowledgments}
		We thank Yun-Tse Tsai for comments on the draft and Yue Zhao for helpful discussions.
		The work of J.B.\ is supported by the National Science Foundation under Grant No.\ 2112789. 
		J.B. thanks the Mainz Institute for Theoretical Physics (MITP) of the Cluster of Excellence PRISMA+ (project ID 39083149) for their hospitality.
	\end{acknowledgments}

	\bibliography{Refs}

	\clearpage
	
	\onecolumngrid
\begin{center}
  \textbf{\large Supplementary Material for Dark Matter Induced  Nucleon Decay Signals in Mesogenesis}\\[.2cm]
  \vspace{0.05in}
  {Joshua Berger, \ and Gilly Elor}
\end{center}

	\section{Signal Generation}

	Signal for this study was generated using a custom fork of the \texttt{GENIE}~\cite{Andreopoulos:2009rq,Andreopoulos:2015wxa} software package.  This fork can be obtained at \texttt{ https://github.com/jberger7/Generator-IND}.  Alternatively, the authors may be contacted for assistance with signal generation.
	
	\begin{table}[!tbh]
		\centering
		\begin{tabular}{|c |c|}
			\hline
			Process & Code \\
			\hline
			$\phi + p \to \xi + \pi^+$ & 61 \\
			$\phi + n \to \xi + \pi^0$ & 62\\
			$\phi + p \to \xi + K^+$ & 63 \\
			$\phi + n \to \xi + K_S^0$ & 64 \\
			$\phi + n \to \xi + K_L^0$ & 65 \\
			$\phi + n \to \xi + D_0$ & 66 \\
			\hline
		\end{tabular}
		\caption{Process codes for IND in the custom fork of \texttt{GENIE} used in this work.}\label{tab:genie-modes}
	\end{table}
	To generate induced nucleon decay signals, the \texttt{GENIE} code should be built following the standard instructions, with the option \texttt{--enable-nucleon-decay}.  The IND processes can be generated by running the standard spontaneous nucleon decay app, \texttt{gevgen\_ndcy}, with the nucleon decay modes 61 through 66.  These correspond to IND processes as indicated in Table \ref{tab:genie-modes}.  In all cases, $\phi$ is assumed to be the incoming dark matter, though nothing in our simulation depends on whether the incoming dark matter is $\phi$ or $\xi$.  Thus, assigning the $\xi$ mass to $\phi$ and vice versa allows for generation of $\xi$ scattering.  The masses of the particles are set in \texttt{config/NucleonDecayPrimaryVtxGenerator.xml}.

	\section{Detector Model}
	
	We consider three types of detectors in this study.  For the purposes of this study, we model detector effects for each in a simplified manner, implementing a kinetic energy threshold on detectable particles and a flat efficiency.  This procedure is highly simplified, but provides a good estimate of the effects of detector and reconstruction efficiencies. to determine the thresholds for meta-stable particles, we combine published results with arguments based on physics limitations when such results are not available.  For JUNO, the thresholds should be very low.  We assume the quoted detection threshold of $0.5~\text{MeV}$~\cite{JUNO:2015sjr} for all meta-stable charged particles, as well as photons.  For Super-Kamiokande and Hyper-Kamiokande, we take electron and photon thresholds from Ref.~\cite{Super-Kamiokande:2016yck}.   For the remaining meta-stable charged particles in the water Cherenkov detectors, we take the threshold to be the Cherenkov threshold in water.  For electrons and photons in LArTPCs, we take the threshold to be the energy required to travel at least one radiation length (roughly $14~\text{cm)}$, such that a shower will develop.  For the remaining track-like particles, we take the threshold to be roughly the energy required to cross at least 10 wires at a wire pitch of $4~\text{mm}$~\cite{PV}, perpendicular to the wires.  In other words, we require that the particles travel a distance of at least $4~\text{cm}$.

	For the efficiencies, we rely on studies of spontaneous proton decay.  For JUNO, only the $K^+ \overline{\nu}$ channel was considered in Ref.~\cite{JUNO:2015sjr}.  A combined branching fraction times efficiency of $0.55$ is found in that channel.  We adopt this efficiency for both the $K^0_S$ and $K^+$ induced decay channels, as well as background that match the signal topology up to thresholds.  The muon reconstruction efficiency is found to be near $1$ elsewhere in the JUNO Conceptual Design Report, so we assume that similar $\pi^+$ topologies have an efficiency of $1$.  We are unaware of a detailed study of $\pi^0$ reconstruction efficiency, so we adopt an efficiency of $1$ for that channel as well, which is likely to be overly optimistic. For Super-Kamiokande and Hyper-Kamiokande, we adopt the average signal efficiencies from Ref.~\cite{Super-Kamiokande:2013rwg,Super-Kamiokande:2014otb} for pion and $K^+$ signals and background candidate events.  For $K^0_S$, we are not aware of any detailed study, so we adopt the same efficiency as for $K^+$.  Finally, for DUNE, we use the efficiencies from the corresponding channel in Tables 7 and 8 of Ref.~\cite{Bueno:2007um} where available.  For $K^0_S$, we note very high efficiencies for $K^0_S e^+/\mu^+$ and so adopt an efficiency of $1$.
	
	In addition to the above assumptions, we need to contend with $\mu^\pm$--$\pi^\pm$ separation.  	In Super-Kamiokande's search for proton decay in the $\pi^+ \nu$ channel, no muon/pion separation was attempted~\cite{Super-Kamiokande:2013rwg}.  JUNO also likely has little capability to distinguish muons from charged pions~\cite{JUNO:2021tll}.  On the other hand, the LArTPC experiment MicroBooNE has demonstrated muon/pion separation at the 80\% level~\cite{Esquivel:2018fls}, consistent with other particle identification efficiencies.  We therefore assume no muon-charged pion separation for water Cherenkov detectors or JUNO, while we assume perfect separation at DUNE for simplicity.  With these assumptions, we find comparable background numbers to our reference searches~\cite{JUNO:2015sjr,Super-Kamiokande:2013rwg,Super-Kamiokande:2014otb}.

	\begin{table}[t!]
		\centering
		\small
		\renewcommand{\arraystretch}{1.1}
		\setlength{\arrayrulewidth}{.25mm}
		\begin{tabular}{|c||c|c|}
			\hline
			Particle & LArTPC & Cherenkov  \\
			& Thresh.~[MeV]  & Thresh.~[MeV]  \\
			\hline
			\hline
			$e^{\pm}$ & 30 & 3.5 \\
			$\gamma$ & 30 & 3.5 \\
			$\mu^{\pm}$ & 35 & 55 \\
			$\pi^{\pm}$ & 35 & 72 \\
			$p$ & 80 & 481 \\
			\hline
		\end{tabular}
		\caption{Kinetic energy thresholds for particles assumed in LArTPC (DUNE) and Water Cherenkov (Super- and Hyper-Kamiokande) detectors. The details of these why these thresholds are chosen are found in the text. }
	\end{table}
	
	\section{Detailed Result Tables}

	In this section, we present detailed information about our signal and background selection and rates.  We include tables for DUNE (Table \ref{tab:DUNE-detailed}), Super-Kamiokande (Table \ref{tab:SuperK-detailed}), Hyper-Kamiokande (Table \ref{tab:HyperK-detailed}), and JUNO (Table \ref{tab:JUNO-detailed}).

	\begin{table}[!tbh]
		\centering
		\begin{tabular}{|c |c |c|| c |c |c |c |c|}
			\hline
		Meson & $m_{\phi}$ & $m_{\xi}$ & $\phi$ eff. & $\xi$ eff. & Bkg. & Bkg. eff. & $S/B$ at sensitivity \\
		\hline\hline
	$\pi^+$ &        0.95 & 0.92 &    48.89 &    49.34 &    118 &     0.10 &     0.63 \\
	$\pi^+$ &        2.45 & 1.53 &    40.32 &     0.00 &     14 &     0.01 &     0.80 \\
	$\pi^+$ &        2.38 & 1.60 &    41.24 &     0.00 &    584 &     0.48 &     0.61 \\
	$\pi^+$ &        0.95 & 0.17 &    38.38 &     0.00 &    600 &     0.49 &     0.61 \\
	$\pi^0$ &        0.95 & 0.92 &    51.24 &    50.65 &    140 &     0.11 &     0.62 \\
	$\pi^0$ &        2.45 & 1.53 &    43.38 &     0.00 &     26 &     0.02 &     0.71 \\
	$\pi^0$ &        2.38 & 1.60 &    44.52 &    80.90 &    915 &     0.74 &     0.60 \\
	$\pi^0$ &        0.95 & 0.17 &    40.75 &    25.43 &    923 &     0.75 &     0.60 \\
	$K^+$ &         0.95 & 0.92 &    83.76 &    85.58 &      0 &     0.00 &  \\
	$K^+$ &        2.45 & 1.53 &    73.72 &     0.00 &      0 &     0.00 &  \\
	$K^+$ &       2.20 & 1.80 &    75.83 &    98.18 &      0 &     0.00 &  \\
	$K^+$ &        0.95 & 0.55 &    78.99 &    94.85 &      0 &     0.00 &  \\
	$K^0_S$ &       0.95 & 0.92 &   100.00 &   100.00 &      0 &     0.00 &  \\
	$K^0_S$ &       2.45 & 1.53 &   100.00 &     0.00 &      0 &     0.00 &  \\
	$K^0_S$ &       2.20 & 1.80 &   100.00 &    97.58 &      0 &     0.00 &  \\
	$K^0_S$ &       0.95 & 0.55 &   100.00 &    93.50 &      0 &     0.00 &  \\
	\hline
\end{tabular}
\caption{Detailed signal and background efficiencies, rates, and signal to background ratios at our estimated sensitivity for DUNE.  For the kaon channels, the background is assumed to be 0, so we do not list $S/B$.}\label{tab:DUNE-detailed}
\end{table}

	\begin{table}[!tbh]
	\centering
	\begin{tabular}{|c |c |c|| c |c |c |c |c|}
		\hline
		Meson & $m_{\phi}$ & $m_{\xi}$ & $\phi$ eff. & $\xi$ eff. & Bkg. & Bkg. eff. & $S/B$ at sensitivity \\
		\hline\hline
  $\pi^+$ &        0.95 & 0.92 &    36.15 &    36.24 &    914 &     1.13 &     0.60 \\
$\pi^+$ &        2.45 & 1.53 &    32.51 &     0.00 &    225 &     0.28 &     0.61 \\
$\pi^+$ &        2.38 & 1.60 &    32.96 &     0.00 &   1336 &     1.65 &     0.60 \\
$\pi^+$ &        0.95 & 0.17 &    32.56 &     0.00 &   1512 &     1.86 &     0.60 \\
$\pi^0$ &        0.95 & 0.92 &    50.96 &    50.15 &    100 &     0.12 &     0.63 \\
$\pi^0$ &        2.45 & 1.53 &    43.58 &     0.00 &     14 &     0.02 &     0.81 \\
$\pi^0$ &        2.38 & 1.60 &    46.06 &    45.86 &    472 &     0.58 &     0.61 \\
$\pi^0$ &        0.95 & 0.17 &    43.82 &    14.64 &    486 &     0.60 &     0.61 \\
$K^+$ &        0.95 & 0.92 &    44.52 &    44.52 &      0 &     0.00 &  \\
$K^+$ &        2.45 & 1.53 &    43.34 &     0.00 &      0 &     0.00 &  \\
$K^+$ &        2.20 & 1.80 &    44.15 &    43.94 &      0 &     0.00 &  \\
$K^+$&        0.95 & 0.55 &    44.52 &    43.25 &      0 &     0.00 &  \\
$K^0_S$ &        0.95 & 0.92 &   100.00 &   100.00 &      0 &     0.00 &  \\
$K^0_S$ &        2.45 & 1.53 &   100.00 &     0.00 &      0 &     0.00 &  \\
$K^0_S$ &        2.20 & 1.80 &   100.00 &    97.05 &      0 &     0.00 &  \\
$K^0_S$ &        0.95 & 0.55 &   100.00 &    94.49 &      0 &     0.00 &  \\
		\hline
	\end{tabular}
	\caption{Detailed signal and background efficiencies, rates, and signal to background ratios at our estimated sensitivity for Super-Kamiokande.  For the kaon channels, the background is assumed to be 0, so we do not list $S/B$.}\label{tab:SuperK-detailed}
\end{table}

	\begin{table}[!tbh]
	\centering
	\begin{tabular}{|c |c |c|| c |c |c |c |c|}
		\hline
		Meson & $m_{\phi}$ & $m_{\xi}$ & $\phi$ eff. & $\xi$ eff. & Bkg. & Bkg. eff. & $S/B$ at sensitivity \\
		\hline\hline
  $\pi^+$ &        0.95 & 0.92 &    36.15 &    36.24 &   4952 &     1.13 &     0.60 \\
$\pi^+$ &        2.45 & 1.53 &    32.51 &     0.00 &   1217 &     0.28 &     0.60 \\
$\pi^+$ &        2.38 & 1.60 &    32.96 &     0.00 &   7237 &     1.65 &     0.60 \\
$\pi^+$ &        0.95 & 0.17 &    32.56 &     0.00 &   8188 &     1.86 &     0.60 \\
$\pi^0$ &        0.95 & 0.92 &    50.96 &    50.15 &    542 &     0.12 &     0.61 \\
$\pi^0$ &        2.45 & 1.53 &    43.58 &     0.00 &     75 &     0.02 &     0.64 \\
$\pi^0$ &        2.38 & 1.60 &    46.06 &    45.86 &   2555 &     0.58 &     0.60 \\
$\pi^0$ &        0.95 & 0.17 &    43.82 &    14.64 &   2634 &     0.60 &     0.60 \\
$K^+$ &        0.95 & 0.92 &    44.52 &    44.52 &      0 &     0.00 &  \\
$K^+$ &        2.45 & 1.53 &    43.34 &     0.00 &      0 &     0.00 &  \\
$K^+$ &        2.20 & 1.80 &    44.15 &    43.94 &      0 &     0.00 &  \\
$K^+$ &        0.95 & 0.55 &    44.52 &    43.25 &      0 &     0.00 &  \\
$K_0^S$ &        0.95 & 0.92 &   100.00 &   100.00 &      0 &     0.00 &  \\
$K_0^S$ &        2.45 & 1.53 &   100.00 &     0.00 &      0 &     0.00 &  \\
$K_0^S$ &        2.20 & 1.80 &   100.00 &    97.05 &      0 &     0.00 &  \\
$K_0^S$ &        0.95 & 0.55 &   100.00 &    94.49 &      0 &     0.00 &  \\
		\hline
	\end{tabular}
	\caption{Detailed signal and background efficiencies, rates, and signal to background ratios at our estimated sensitivity for Hyper-Kamiokande.  For the kaon channels, the background is assumed to be 0, so we do not list $S/B$.}\label{tab:HyperK-detailed}
\end{table}

	\begin{table}[!tbh]
	\centering
	\begin{tabular}{|c |c |c|| c |c |c |c |c|}
		\hline
		Meson & $m_{\phi}$ & $m_{\xi}$ & $\phi$ eff. & $\xi$ eff. & Bkg. & Bkg. eff. & $S/B$ at sensitivity \\
		\hline\hline
  $\pi^+$ &        0.95 & 0.92 &    71.26 &    72.11 &     53 &     0.11 &     0.66 \\
$\pi^+$ &        2.45 & 1.53 &    65.15 &     0.00 &      8 &     0.02 &     0.94 \\
$\pi^+$ &        2.38 & 1.60 &    66.51 &    29.11 &    166 &     0.35 &     0.62 \\
$\pi^+$ &        0.95 & 0.17 &    64.93 &    23.82 &    177 &     0.38 &     0.62 \\
$\pi^0$ &        0.95 & 0.92 &    63.88 &    63.18 &     50 &     0.11 &     0.66 \\
$\pi^0$ &        2.45 & 1.53 &    57.99 &     0.00 &      8 &     0.02 &     0.91 \\
$\pi^0$ &        2.38 & 1.60 &    59.70 &    30.43 &    148 &     0.32 &     0.62 \\
$\pi^0$ &        0.95 & 0.17 &    57.55 &    10.80 &    159 &     0.34 &     0.62 \\
$K^+$ &        0.95 & 0.92 &    59.15 &    59.25 &      0 &     0.00 &  \\
$K^+$ &        2.45 & 1.53 &    57.75 &     0.00 &      0 &     0.00 &  \\
$K^+$ &        2.20 & 1.80 &    58.45 &    60.41 &      0 &     0.00 &  \\
$K^+$ &        0.95 & 0.55 &    59.25 &    59.98 &      0 &     0.00 &  \\
$K_0^S$ &        0.95 & 0.92 &    55.00 &    55.00 &      0 &     0.00 &  \\
$K_0^S$ &        2.45 & 1.53 &    55.00 &     0.00 &      0 &     0.00 &  \\
$K_0^S$ &        2.20 & 1.80 &    55.00 &    53.12 &      0 &     0.00 &  \\
$K_0^S$ &        0.95 & 0.55 &    55.00 &    51.54 &      0 &     0.00 &  \\
		\hline
	\end{tabular}
	\caption{Detailed signal and background efficiencies, rates, and signal to background ratios at our estimated sensitivity for JUNO.  For the kaon channels, the background is assumed to be 0, so we do not list $S/B$. Efficiencies  are taken from \cite{JUNO:2015sjr}. Note that a more recent proton decay search at JUNO used slightly smaller efficiency \cite{JUNO:2022qgr}, and very different kinematic cuts than are needed for the IND signal. We hope that this work motivates the JUNO collaboration to do a dedicated IND search.}	
	\label{tab:JUNO-detailed}
\end{table}

\end{document}